\documentclass[preprint,proceedings]{rmaa}

\title{The X-ray view of the ionization cone in NGC~5252}
\author{Matteo Guainazzi\altaffilmark{1},
	Stefano Bianchi\altaffilmark{2},
	Massimo Cappi\altaffilmark{3},
	Mauro Dadina\altaffilmark{3},
	Giuseppe Malaguti\altaffilmark{3}
  }

\suppressfulladdresses

\altaffiltext{1}{European Space Astronomy Center of ESA, Apartado 50727, E-28080 Madrid, Spain; email: \texttt{Matteo.Guainazzi@sciops.esa.int}.}
\altaffiltext{2}{Dipartimento di Fisica ``E.Amaldi'', Universit\`a di Roma Tre, Via della Vasca Navale, I-00146, Roma, Italy.}
\altaffiltext{3}{INAF-IASF Bologna, via Gobetti 101, 40129 Bologna, Italy.}

\shortauthor{Guainazzi et al.} \shorttitle{X-ray ionization cone in NGC~5252}

\resumen{} 

\abstract{We present the results of a {\it Chandra} soft X-ray observation of
the spectacular
ionization cone in the nearby Seyfert~2 galaxy NGC~5252.
As almost invariably observed in obscured AGN, the soft X-ray emission
exhibits a remarkable morphological concidence with the cone
ionized gas as traced
by HST O[{\sc iii}] images. Energy-resolved images and high-resolution
spectroscopy suggest that the X-ray emitting gas is photoionized by the AGN,
at least on scales as large as the innermost gas and stellar ring ($\le$3~kpc).
Assuming that the whole cone is photoionized by the AGN, we reconstruct the
history of the active nucles in the last $\sim$10$^5$~years.}

\begin{document}
\maketitle

The Seyfert~2 galaxy NGC~5252 hosts one of the most spectacular
ionization cones ever observed in an Active Galactic Nucleus (AGN).
Line emission exhibits a
bi-conical morphology \citep{tadhunter89} extending out to $\simeq$20~kpc
either side of the active nucleus. Optical studies have suggested that
O[{\sc iii}] emission is due to photoionization by the active nucleus.
On smaller scales
($\simeq$3~kpc) a gas disk \citep{tsvetanov96} is probably
due to the combination of three dynamical components: an inclined
($\simeq$40$^{\circ}$) stellar disk, and two gas disks associated with it and
rotating in opposite directions. It has been suggested that the gas of this
inner structure is also photoionized by the AGN \citep{morse98}.

Recent {\it Chandra} high-resolution observations 
have revealed large scales (hundreds of parsecs to kilo-parsecs)
soft X-ray emission in almost all nearby obscured AGN
with Extended Narrow Line Regions (ELNRs)
where this measurement is technically possible \citep{bianchi06}.
Soft X-rays show a remarkable morphological coincidence with
high-resolution O[{\sc iii]} HST images. 
High-resolution spectra taken with the Reflection Grating
Spectrometer on-board XMM-Newton
confirm that the soft X-rays carry the
unmistakable signatures of photoionized gas: ``narrow''
($\delta E$$\simeq$1--10~eV) Radiative Recombination Continua,
and large ratios between
the forbidden component of the He-like H$\alpha$ triplets and
the H-like Ly$\alpha$
\citep{sako00,sambruna01,kinkhabwala02,guainazzi07}. The large
intensity of higher-order transitions with respect to the
K$_{\alpha}$ are indicative of an important role played by
resonant scattering.

In this paper, we briefly discuss
the soft X-ray properties
of NGC~5252. At its distance (92~Mpc) the spatial scale is: 1$\arcsec$ = 450~pc.

\begin{figure}
\begin{center}
\includegraphics[angle=-90,width=8.0cm]{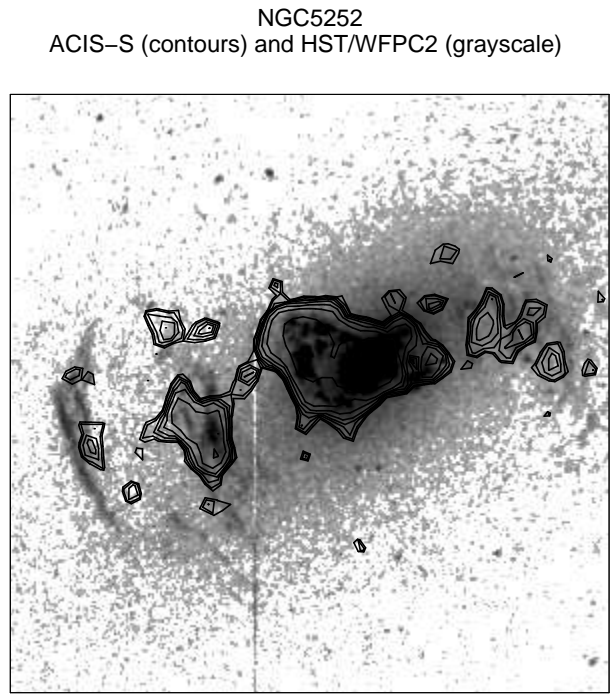}
\end{center}
\caption{O[{\sc iii}] HST/WFC2 ({\it greyscale}) and 0.2-1~keV Chandra/ACIS
({\it smoothed contours}) images of NGC~5252. The image is 35$\arcsec$ aside.}
\label{fig1}
\end{figure}

\begin{figure}
\begin{center}
\includegraphics[width=8.0cm]{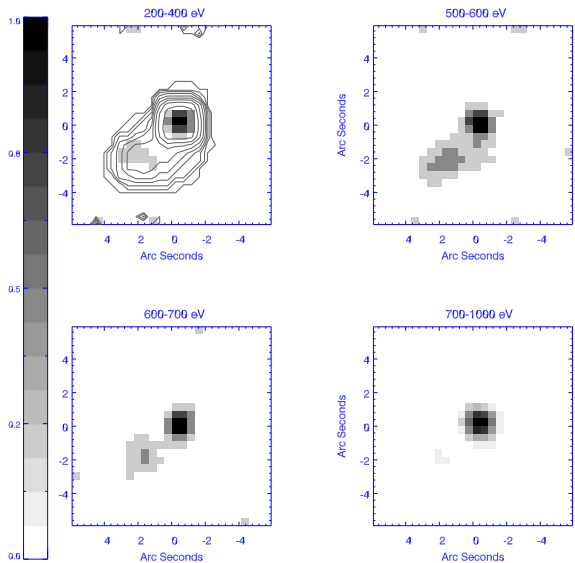}
\end{center}
\caption{Energy resolved {\it Chandra}/ACIS images of NGC~5252. The
coordinate reference system is centered on the position of the X-ray active
nucleus \citep{cappi96}. }
\label{fig2}
\end{figure}

\begin{figure}
\begin{center}
\includegraphics[width=7.0cm]{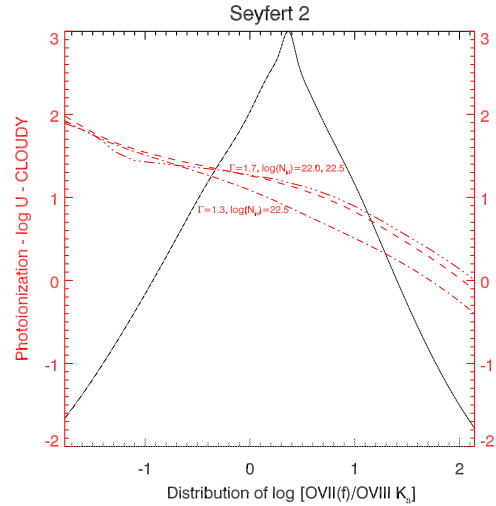}
\end{center}
\caption{{\it Solid line}: distribution of the observed values of the
flux logarithmic ratio between the O{\sc vii} He$\alpha$
forbidden component and the O{\sc viii} Ly$\alpha$;
{\it dashed and dotted lines}: {\sc Cloudy} predictions
as a function of the ionization parameter $U$ (y-axis)
for different values of the gas column density and
ionizing continuum X-ray photon index $\Gamma$.}
\label{fig5}
\end{figure}

The soft X-ray (0.2-1~keV) {\it Chandra}/ACIS
emission in NGC~5252 is extended on scales as
large as $\simeq$11~kpc (Fig.~\ref{fig1}). 
The hard X-ray (1--10~kpc) image is instead unresolved.
There is a
good morphological coincidence between the X-rays and the 
O[{\sc iii}] emission, taking into
account the different spatial resolutions. The agreement extends also
to the diffuse soft X-ray emission overlapping with the smaller scale
($\simeq$3~kpc) stellar and gaseous disks. On this small scale
X-rays are mostly
due to He-like and H-like Oxygen recombination transitions, with a smaller
contribution by C{\sc vi} and by the Fe-L complex (Fig.~\ref{fig2}).
The measured flux ratio between the counts in the O{\sc vii} and
O{\sc viii} bands, once corrected for the energy dependence of the ACIS
effective area ($2.3 \pm 0.4$), is typical of AGN-photoionized sources
\citep{guainazzi07}. This correspond to ionization
parameters $\log(U) \sim 1$
\footnote{$U$ is defined as $\frac{\Phi(H)}{n(H) c}$, where
$\Phi(H)$ is the surface flux of ionizing photons, and
$n(H)$ is the total hydrogen density} (see Fig.~\ref{fig5}).

\begin{figure}[h]
\begin{center}
\includegraphics[width=8.0cm]{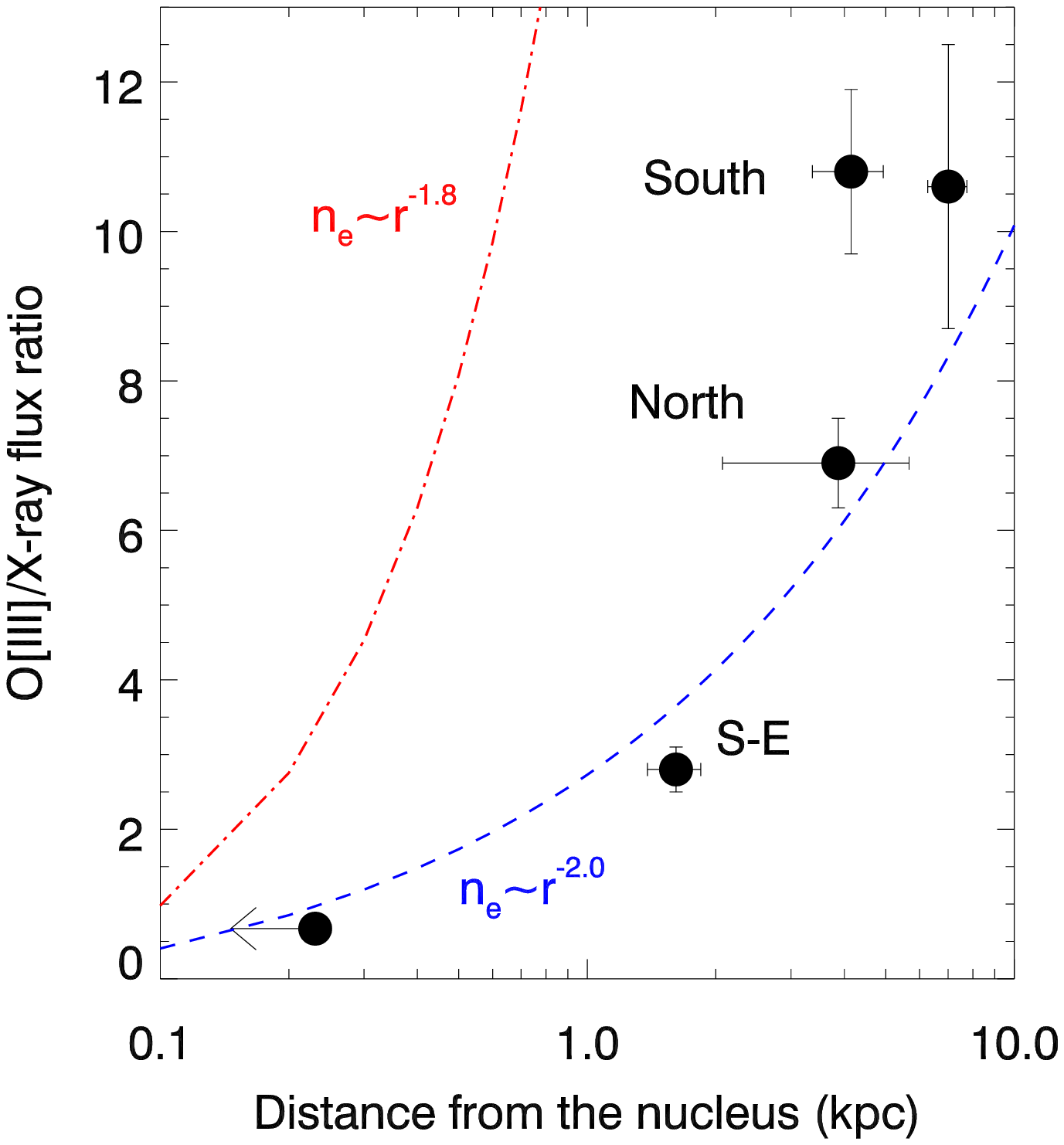}
\end{center}
\caption{O[{\sc iii}] to soft X-ray flux ratio along the ionization cone
(the x-axis is the distance from the optical position of the
active nucleus). The lines represent the predictions of a {\sc CLOUDY}-based
\citep{ferland98} model in photoionization equilibrium,
where the AGN is responsible for photoionizing the
cone gas. The AGN Spectral Energy Distribution has: $\alpha_{ox} = -1.4$,
${\alpha_UV} = -0.50$, and $\alpha_X = 2.0$. An X-ray to bolometric
luminosity
correction of 1/0.03 is assumed \citep{elvis94}.}
\label{fig3}
\end{figure}

In
Fig.~\ref{fig3} we show the O[{\sc iii}] to soft X-ray flux ratio as a
function of
the radius along the ionization cone. The ratio is calculated at the
position of optically bright knots and filaments along the cone.
On the same plot, we compare the predictions of simple, homogeneous and
time-independent photoionization models, following the method described
in Bianchi et al. (2006). An almost constant ionization parameter along
the cone is required, implying a radial decrease of the electronic density as:
$n(r) \propto r^{-(1.8-2.0)}$. Similar trends had been observed in other
obscured AGN with ENLRs \citep{bianchi06}.
Such a decrease is steeper than required by optical diagnostics
of space resolved NLRs (Bennert et al. 2006), and may suggest that a local
photoionizing source contribute to the soft X-ray emission.
Shocks heating of the hot gas by stellar winds or
interaction with a feeble jet are possible culprits.

\begin{figure}
\begin{center}
\includegraphics[width=8.0cm]{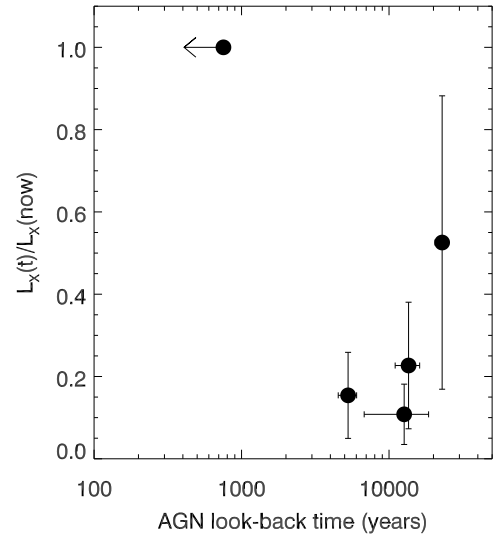}
\end{center}
\caption{History of the ionizing flux responsible for
the ionization cone in NGC~5252. Details in text }
\label{fig6}
\end{figure}

If AGN photoionization is still responsible for the bulk
of the ionization equilibrium in the gas, one can derive
from the results in Fig.~\ref{fig3} and the known
geometry of the cone knots and filaments the history of
the active nucleus responsible for the ionization cone observed
nowadays (Fig.~\ref{fig6}).
Its average level of activity in the last $\sim$1000~years is
a factor of 2--8 larger than when the currently visible ionization cones
were lit-up.

A more detailed description of the observations and of
their implications will
be the subject of a subsequent paper (Guainazzi et al., in preparation).


\begin{thebibliography}

\bibitem[Bennert et al. 2006]{bennert06} Bennert N., Jungwiert B., Komossa S., Haas M., Chini R. 2006, A\&A, 456, 953

\bibitem[Bianchi et al. 2006]{bianchi06} Bianchi S., Guainazzi M., Chiaberge M. 2006, A\&A, 448, 499

\bibitem[Cappi et al. 1996]{cappi96} Cappi M., Mihara T., Matsuoka M., Brinkmann W., Prieto M.A., Palumbo G.G.C. 1996, ApJ, 456, 141
\adjustfinalcols
\bibitem[Elvis et al. 1994]{elvis94} Elvis M., et al. 1994, ApJS, 95, 1

\bibitem[Ferland et al. 1998]{ferland98} Ferland G. J., Korista K.T., Verner D.A., Ferguson J.W., Kingdon J.B., Verner E.M. 1998, PASP, 110, 761

\bibitem[Guainazzi \& Bianchi 2007]{guainazzi07} Guainazzi M., Bianchi S. 2007, 374, 1290

\bibitem[Kinkhabwala et al. 2002]{kinkhabwala02} Kinkhabwala A., et al. 2002, ApJ, 575, 732 

\bibitem[Morse et al. 1998]{morse98} Morse J.A., Cecil G., Wilson A.S., Tsvetanov Z.I. 1998, ApJ, 505, 159

\bibitem[Sako et al. 2000]{sako00} Sako M., Kahn S.M., Paerels F., Liedahl D.A. 2000, ApJL 543, L115

\bibitem[Sambruna et al. 2001]{sambruna01} Sambruna R., et al. 2001, ApJ, 546, L13

\bibitem[Tadhunter \& Tsvetanov 1989]{tadhunter89} Tadhunter C., Tsvetanov Z. 1989, Nature, 341, 422

\bibitem[Tsvetanov et al. 1996]{tsvetanov96} Tsetanov Z.I., Morse J.A., Wilson A.S., Cecil G. 1996, ApJ, 458, 172

\end{thebibliography}
\end{document}